\documentclass{aa}
\usepackage{psfig}
\usepackage{times}

\def\degr{$^{\circ}$}
\begin{document}
\thesaurus{06                  
           (08.09.2 HD 199143; 
            08.16.1;           
            08.18.1;           
            10.15.1;           
            13.21.5)}          
\headnote{Letter to the editor}
\title{A young stellar group associated with HD~199143 ($d$ = 48~pc)
\thanks{Based on observations collected at the European Southern Observatory, 
        La Silla, Chile, observations made by {\it IUE} at NASA-GSFC, and  
        with the Isaac Newton Telescope operated on the island of La Palma by 
        the Isaac Newton Group in the Spanish Observatorio del Roque de los 
        Muchachos}}

\author{M.E. van den Ancker\inst{1,2} \and M.R. P\'erez\inst{3} \and 
        D. de Winter\inst{4,5} \and B. McCollum\inst{6}}
\institute{
Astronomical Institute ``Anton Pannekoek'', University of Amsterdam, 
 Kruislaan 403, 1098 SJ  Amsterdam, The Netherlands \and
Harvard-Smithsonian Center for Astrophysics, 60 Garden Street, MS 42, 
 Cambridge  MA 02138, USA \and
Emergent-IT Corp., 9315 Largo Drive West, Suite 250, Largo  MD 20774, USA \and 
Instituto de Astrof\'{\i}sica de Canarias, C/ Via L\'{a}ctea s/n, 
 38200 La Laguna, Tenerife, Spain \and
TNO-TPD, Stieltjesweg 1, P.O. Box 155, 2600 AD  Delft, The Netherlands \and
IPAC-Caltech, SIRTF Science Center, MS 314-6, Pasadena, CA 911125, USA}

\offprints{M.E. van den Ancker (mario@astro.uva.nl)}
\date{Received <date>; accepted <date>} 

\maketitle

\begin{abstract}
We present new optical and ultraviolet spectroscopy of the anomalous 
EUV emitter HD~199143 (F8V). High resolution spectra in the H$\alpha$ and 
Na\,{\sc i}\,D wavelength regions show evidence for very rapid (a few 
hundred km~s$^{-1}$) rotation of the stellar photosphere. Using archive 
{\it IRAS} data we also show that the star has excess emission above 
photospheric levels at 12~$\mu$m. {\it IUE} data of HD~199143 reveal the 
presence of emission lines of Mg\,{\sc ii}, C\,{\sc i}, C\,{\sc ii}, 
C\,{\sc iii}, C\,{\sc iv}, Si\,{\sc iv}, He\,{\sc ii} and N\,{\sc v} and 
show a large variability, both in the continuum and line fluxes.  
We propose that all available data of HD~199143 can be explained by 
assuming that is has been spun up by accretion of material from a
close T~Tauri like companion, responsible for the emission lines, 
the ultraviolet variability and the excess infrared emission. 
The bursting or flaring nature of this object, mostly in high energies, 
could be explained as episodic mass transfer between the star and its 
close companion. 
We show that HD~199143 and the Li-rich late-type dwarf BD$-$17\degr6128 
form a physical pair and suggest that both may be part of a new nearby 
(48~pc) young ($\sim$ 10$^7$ yr) stellar association in Capricornius. 
\end{abstract}

\section{Introduction}
Zuckerman \& Webb (2000) sketch a picture of the recent star 
formation history of the solar neighbourhood in which 
10--40 million years ago an ensemble of molecular clouds 
were forming stars at a modest rate near the present position 
of the Sun. About 10~Myrs ago, the most massive of these 
newly formed stars exploded as a supernova, terminating the 
star formation episode and generating the very low density 
region seen in most directions from the present Sun (Welsh 
et al. 1998). This scenario can not only explain the presence 
of young stellar groups close to the earth, but also explains 
how the $\beta$~Pic moving group can be so young (20~Myr; 
Barrado y Navascu\'es et al. 1999), and yet so close. However, 
currently this scenario is largely speculative.

HD~199143 is a poorly studied bright ($V$ = 7\fm27) star in the 
constellation of Capricornius. It has been classified as F8V  
in the Michigan Spectral Survey (Houk \& Smith-Moore 1988), after 
an initial classification of G0 by Cannon \& Mayall (1949). The 
star would be completely inconspicuous, if it hadn't been detected 
as a bright extreme-ultraviolet source by the {\it ROSAT} 
(2RE J205547$-$170622) and {\it Extreme Ultraviolet Explorer} 
(2EUVE J2055$-$17.1) missions (Pounds et al. 1993; 
Malina et al. 1994; Pye et al. 1995; Bowyer et al. 1996).

In this {\it Letter} we present new optical and ultraviolet 
spectroscopy of HD~199143 and show that it is a variable and rapidly 
rotating F8V star. We infer that all characteristics of the HD~199143 
system can be explained by assuming that it is a binary system, 
in which the primary has been spun up by accretion of mass from a 
low-mass companion. Its association with a previously studied 
T~Tauri-like system (BD$-$17\degr6128) suggests that these two stars 
could be the first two members of a close (48~pc) new region of 
recent star formation and may provide compelling support for the 
star formation history of the solar neighbourhood outlined in 
the first paragraph.

\section{Optical Observations}
\begin{figure}
\centerline{\psfig{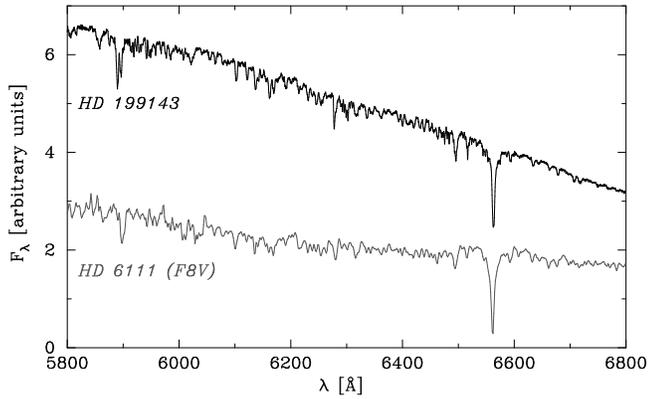}}
\caption[]{Red spectrum of HD 199143 obtained at the INT. For comparison 
we also show the spectrum of HD 6111 (F8V) from the spectral database 
by Jacoby et al. (1984).}
\end{figure}
Low-resolution (0.53~\AA~pix$^{-1}$) spectra of HD~199143 in the 
wavelength region of 5700--6800~\AA\ were taken with the 2.5~m 
Isaac Newton Telescope (INT) at La Palma during the nights of August 
29 (JD~2451023.072), 30 (JD~2451024.122) and 31 (JD~2451025.076), 
1998. The spectra were reduced with the usual steps of bias subtraction, 
flatfielding, background subtraction and spectral extraction, and 
wavelength and flux calibration. Apart from a multiplicative factor, 
due to the fact that the nights in which the spectra were taken 
were not of photometric quality, the spectra taken in the different 
nights were identical. In Fig.~1 we show the spectrum of Aug. 30, 
1998. For comparison we also show the spectrum of HD~6111 (F8V), 
obtained from the spectral database by Jacoby et al. (1984). The 
resolution of this spectrum is about three times lower than that 
of the INT spectrum. Apart from the differences expected because 
of the differences in spectral resolution, the two spectra are 
identical, confirming the F8V spectral classification of HD~199143.

High-resolution (0.05~\AA~pix$^{-1}$) spectra of HD~199143 in 
the H$\alpha$ (6536--6591~\AA) and Na\,{\sc i}\,D (5858--5910~\AA) 
wavelength ranges were obtained with the Coud\'e Auxiliary 
Telescope (CAT) at La Silla, Chile, on Dec. 16 
(at JD~2450432.016) and Dec. 15 (JD~2450431.016), 1996. The 
spectra were reduced in a standard fashion, after which the 
continuum was normalized to unity. They are shown in Figs.~2 
and 3, together with the spectra of HR~963 
(F8V), obtained during the same night as the HD~199143 
spectra. Apart from a number of very sharp 
absorption features due to water vapour in the earth's 
atmosphere, a number of highly broadened 
(FWHM $\approx$ 250~km~s$^{-1}$) photospheric absorption 
lines (most prominently H$\alpha$ and the Na\,{\sc i} 
doublet) are visible in the HD~199143 spectra. The same lines 
are present in HR~963, but much narrower, again confirming 
the spectral classification of F8. From the photospheric 
lines we measure a radial velocity of $-$9 $\pm$ 16 km~s$^{-1}$ 
for HD~199143.  The wings of the H$\alpha$ 
profile appear identical in HD~199143 and HR~963, demonstrating 
that the broadening of the lines in HD~199143 is not due to 
a luminosity classification smaller than V, but must be caused 
by a high (a few hundred km~s$^{-1}$) value of $v \sin i$.
\begin{figure}
\centerline{\psfig{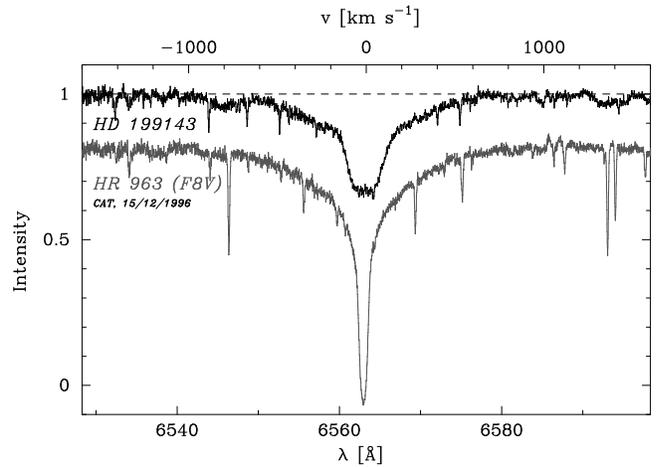}}
\caption[]{High-resolution spectrum of HD 199143 in the H$\alpha$ 
wavelength region. For comparison we also show the spectrum of 
HR~963 (F8V), shifted for clarity, obtained during the same night.}
\end{figure}
\begin{figure}
\centerline{\psfig{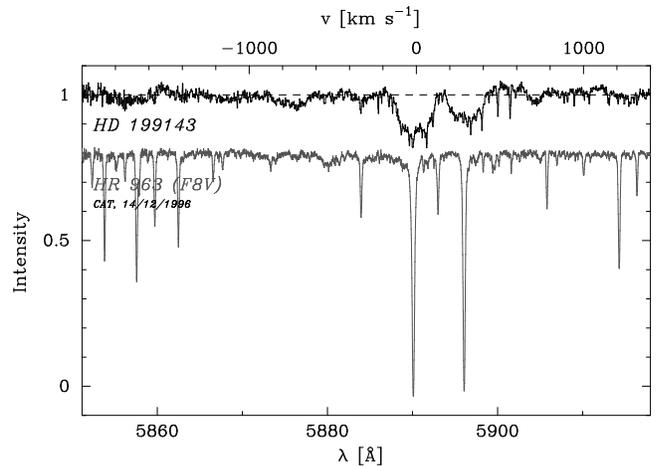}}
\caption[]{High-resolution spectrum of HD 199143 in the Na\,{\sc i}\,D  
wavelength region. We again show the spectrum of HR~963 (F8V) for 
comparison.}
\end{figure}

\section{Ultraviolet Observations}
HD~199143 was observed with {\it IUE} on several occasions in 1995, under a 
discretionary program (OD89Z), and the observatory program (USSBS). Archival data taken 
by the NC119 program, in November 1992, are also included in our analysis. A total of 
19 archived images are available, both in low- (1.68~\AA~pix$^{-1}$ for SWP, 
2.66~\AA~pix$^{-1}$ for LWP) and high-dispersion (25 km s$^{-1}$, 
$\lambda~\sim~0.2$~\AA\ resolution), which were secured through the large aperture 
(oval-shaped: 10\arcsec\ $\times$ 20\arcsec). There are 11 SWP camera (1150--2000~\AA) 
exposures all in low-dispersion, and eight LWP (1800--3200~\AA) exposures, of which 
six are in high-dispersion.

A preliminary inspection of the short-wavelength data reveals an emission spectrum 
typical of T Tauri, Herbig Ae/Be (HAeBe) stars and planetary nebulae.  The SWP 
low-dispersion images present clear emissions in N\,{\sc v} (1240 \AA), 
C\,{\sc ii} (1335 \AA), Si\,{\sc iv} (1394 \AA), C\,{\sc iv} (1550 \AA), 
C\,{\sc iii} (1577 \AA), He\,{\sc ii} (1640 \AA) and C\,{\sc i} (1657 \AA). 
With the exception of a few lines such as N\,{\sc v} and He\,{\sc ii}, the 
emission spectrum of HD 199143 resembles the spectra of T Tauri stars such as 
RW Aur and GW Ori (Imhoff \& Appenzeller 1987). In Fig.~4 we present the emission 
spectrum corresponding to the well-exposed (80 min.) image SWP 55183. This spectrum is 
compared with the spectrum of HR 963 (SWP 52921), which is a typical F8V, 
high-proper motion star.
\begin{figure}
\centerline{\psfig{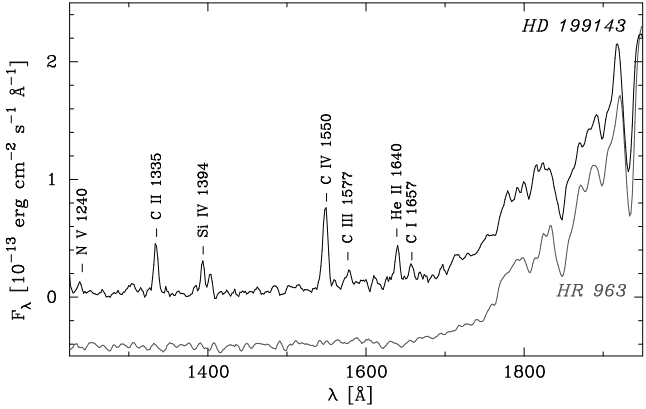}}
\caption[]{{\it IUE} Short wavelength spectrum of HD 199143 (top) 
identifying the emission lines. For comparison we also show the shifted spectrum of 
the bright F8V star HR 963 (bottom), which is clearly devoid of emission lines.}
\end{figure}
\begin{figure}[t]
\centerline{\psfig{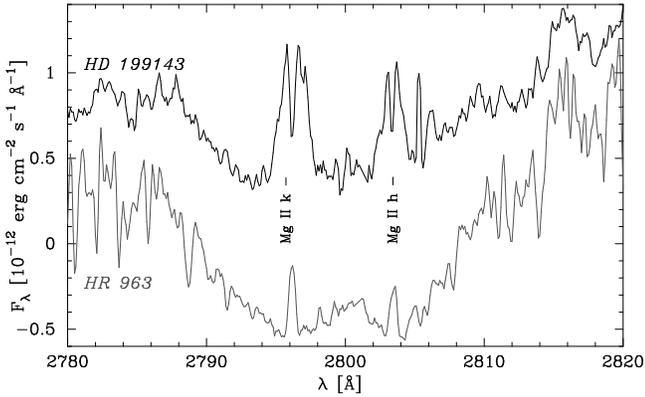}}
\caption[]{{\it IUE} Long wavelength high-resolution spectrum of HD 199143 (top) and 
HR~963 (bottom) centered around the 2800~\AA\ Mg\,{\sc ii} doublet (h \& k). The broad 
photospheric absorption at these wavelengths is also clearly identifiable.  Broad 
circumstellar emission can be seen in the case of HD~199143, whereas incipient 
Mg\,{\sc ii} emission lines are detected in HR~963, typical of late-type main 
sequence stars showing an onset of chromospheric activity.}
\end{figure}

In the long-wavelength spectrum the most conspicuous emission is the Mg\,{\sc ii} doublet 
at 2798.0 and 2802.7 \AA, which is only detectable in high-dispersion exposures. This 
normally P-Cygni emission has been shown to be typical of chromospherically active stars 
and other pre-main sequence (PMS) stars such as T Tauri and HAeBe objects. 
The Mg\,{\sc ii} doublet from the well-exposed (110 min.) image LWP 30968 is 
presented in Fig. 5 along with the comparison spectrum of HR 963.  Note the 
self-absorption feature on top of the h \& k emissions.  The Mg\,{\sc ii} spectrum 
of HD~199143 resembles that of the T Tauri star GW Ori (Imhoff \& Appenzeller 1987), 
or that of the ``double emission peak'' HAeBe stars classification described 
by Imhoff (1994).

\section{Analysis}
The ultraviolet emission spectrum of HD 199143 is somewhat peculiar because emission 
lines such as N\,{\sc v}, which corresponds to a temperature regime of about 
$2 \times 10^{5}$ K, is rarely present in PMS stars. Furthermore, lines such as 
He\,{\sc ii}, which has a complex origin, are commonly present in planetary nebulae 
and in only a handful young objects. The ratio of the emission fluxes for C\,{\sc iv} 
to Si\,{\sc iv} in T Tauri stars ranges from 2--3 (typical of chromospheric activity) 
to less than unity (Imhoff \& Appenzeller 1987).  In HD~199143 this range is 
from 1.2 to 1.4, compatible with an origin in a T Tauri star.

Repeated observations with {\it IUE} allow us to address the issue of ultraviolet 
variability, beyond the instrument flux repeatability (3\%). Comparisons of the 
well-exposed section of the {\it IUE} images indicate a variability of 10--20\% in 
the continuum and 20-50\% in the emission lines of C\,{\sc ii}, C\,{\sc iv}, 
N\,{\sc v} and He\,{\sc ii}. From the repeated SWP images taken closely in 
time, we detected that flux variability (continuum and lines) was found to be 
random and not associated with the period of 1.6 days recently suggested by 
Handler (1999), whom classified HD 199143 as a $\gamma$ Doradus candidate. 

We have searched for infrared emission by checking the raw {\it IRAS} scans 
at the position of HD~199143 using routines from the Groningen Image 
Processing System ({\sc GIPSY}). In the 
12~$\mu$m band, a point-like source is clearly present at the position of 
HD~199143. No source was detected at longer wavelengths. From these data 
we derive a flux of 0.24 $\pm$ 0.04~Jy in the {\it IRAS} 12~$\mu$m band, 
and upper limits of 0.12, 0.12 and 0.30~Jy for the fluxes at 25, 60 and 
100~$\mu$m, respectively.

Using the newly determined {\it IRAS} fluxes and the optical photometry of 
HD~199143 by Olsen (1983) and Cutispoto et al. (1999), we constructed a 
Spectral Energy Distribution (SED) of HD~199143, shown in Fig.~6. In this 
plot we also show UV fluxes of HD~199143 from archive {\it IUE} data. 
Also plotted is a Kurucz (1991) model for the photosphere of a F8V star, 
fitted to extinction-corrected optical photometry of HD~199143. In the 
SED we can see that both the {\it IRAS} 12~$\mu$m flux and the {\it IUE} 
fluxes below 2250~\AA\ are significantly higher than that expected from 
the stellar photosphere. One explanation 
for the infrared excess could be the presence of circumstellar dust 
in the system, similar to that found in Vega-type systems. However, 
the tight 25 and 60~$\mu$m upper limits show that if this is the case, 
only a very warm ($>$ 1000~K) dust component must be present, which 
is unlikely.  A more likely source of the observed infrared excess 
might be a late-type companion to HD~199143, or to infer a modification 
of the photospheric structure of HD~199143 due to its rapid 
rotation (Sect.~2).
\begin{figure}[t]
\vspace{0.3cm}
\centerline{\psfig{figure=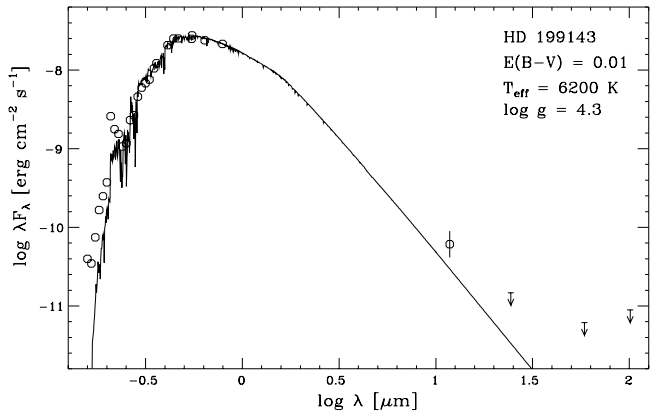,width=8.5cm,angle=90}}
\caption[]{Observed spectral energy distribution of HD 199143 (circles) 
compared to a Kurucz (1991) model for $T_{\rm eff}$ = 6,200~K and 
$\log g$ = 4.3 (solid line).}
\end{figure}

The {\it Hipparcos} catalogue (ESA 1997) lists a parallax of 
21.0 $\pm$ 1.0 milliarcseconds for HD~199143 (HIP~103311). Using a 
distance $d$ of 47.7 $\pm$ 2.4~pc obtained from this parallax
we computed the luminosity of HD~199143 by integrating the flux of the 
Kurucz (1991) stellar photosphere model fitted to the optical photometry, 
and multiplying by $4 \pi d^2$ to correct for spatial dilution. Note that 
if the photospheric structure of HD~199143 has been altered by its rapid 
rotation, this procedure might not be completely 
correct since the emitted flux will be non-spherically symmetric 
distributed. However, the resulting stellar luminosity of 
2.4 $\pm$ 0.2~L$_\odot$ is in agreement with that expected of 
a F8V star (2.1~L$_\odot$; Schmidt-Kaler 1982), showing that this is 
not a big effect. It also confirms our earlier conclusion that 
HD~199143 belongs to luminosity class V.

\section{Discussion and Conclusions}
The presence of a normal late-type companion could not explain the 
ultraviolet excess, or the {\it EUVE} and {\it ROSAT} detections 
of HD~199143. However, the presence of an accretion disk around 
our hypothetical companion, such as that found in LMXB or T~Tauri 
systems, might easily explain those properties, as well as the infrared 
excess, the presence of emission lines and the variability. In such 
a scenario, the high rotational velocity of HD~199143 could be due 
to a spin-up in its past by accretion from the companion. 

At first glance, a scenario in which a nearby main-sequence star like 
HD~199143 would have a T~Tauri-like companion would seem far-fetched. 
However, Mathioudakis et al. (1995) report the presence a strongly 
flaring K7e--M0e dwarf with a high Li abundance only 5 arcminutes 
from HD~199143. The optical spectrum of this star, BD$-$17\degr6128, 
is identical to that of many T~Tauri stars.  
From Digital Sky Survey images we identify BD$-$17\degr6128 with 
HD~358623. An inspection of the Tycho-2 Catalogue (H{\o}g et al. 
2000) shows that this star has a proper motion of 59 $\pm$ 3 and 
$-$63 $\pm$ 3 mas~yr$^{-1}$ in $\mu_\alpha$ and $\mu_\delta$, 
identical to that of HD~199143. From the fact that HD~358623 is 
the only star within a 5 degree radius for which this is the case, 
we exclude the possibility that this could be a coincidence and 
conclude that the two stars form a genuine proper motion pair. 
Using the data by Mathioudakis et al. (1995), and the newly 
determined distance, we compute the absolute luminosity of 
BD$-$17\degr6128 to be 0.34 $\pm$ 0.06 L$_\odot$, employing a 
similar procedure to that followed for HD~199143. Comparison 
with the pre-main sequence evolutionary tracks by 
D'Antona \& Mazzitelli (1997) yields an age of 10$^7$ 
years for BD$-$17\degr6128, consistent with a T~Tauri 
nature of this star.  

Using the radial velocity of HD~199143 determined in Section~2, 
and the parallax and proper motions listed in the {\it Hipparcos} 
catalogue, we compute the galactic space velocity 
components $(U, V, W)$ of HD~199143 to be 
$(-10 \pm 13, -13 \pm 6, -13 \pm 6)$ km~s$^{-1}$.  This space 
motion is similar to that of many stars in the vicinity 
of the Tucanae and TW Hydra associations (Zuckerman \& Webb 2000), 
suggesting that these stars might have formed from the same 
cloud complex.  We conclude that HD~199143 and BD$-$17\degr6128
could very well be the first two members of a region of 
recent star formation similar to the TW Hydrae Association and the 
newly identified Tucanae Association (Kastner et al. 1997; 
Zuckerman \& Webb 2000). If confirmed, a further study of these 
two enigmatic stars could lead to a better understanding of 
the star formation history in the solar neighbourhood.

\acknowledgements{The authors would like to thank the referee, 
Ben Zuckerman, for valuable suggestions for improving the 
manuscript.  Bruce McCollum and Mario P\'erez thank Yoji Kondo for 
Discretionary Time to observe HD~199143 with {\it IUE}. 
This research has 
made use of the Simbad data base, operated at CDS, Strasbourg, France.}

\end{document}